# AXIONS IN ASTROPHYSICS AND COSMOLOGY[†]

Georg G. Raffelt

Max-Planck-Institut für Physik

Föhringer Ring 6, 80805 München, Germany

## ABSTRACT

If axions exist they are efficiently produced in the hot and dense interior of stars, providing a novel energy-loss mechanism. In order to avoid a conflict with the observed properties of stars, one can derive a lower limit on the Peccei-Quinn scale (an upper limit on the axion mass). In the early universe, axions are produced by the "misalignment mechanism" and the emission from global strings as well as the relaxation of the string-domain wall system formed at the QCD phase transition. In order to avoid an "overclosed universe" the Peccei-Quinn scale must obey an upper limit (a lower limit on the axion mass). The current values of these bounds are reviewed. There remains a "window of opportunity" $10^{-5}\,\text{eV} \lesssim m_a \lesssim 10^{-2}\,\text{eV}$, with large uncertainties on either side, where axions could still exist.

---

I. INTRODUCTION

The interest in axions as a possible dark matter candidate has recently soared thanks to the heroic progress made by the two groups who are currently mounting new search experiments which have a realistic chance of finding these elusive particles if they are the dark matter of the galaxy[1]. For such efforts to make sense one needs to understand the "window of opportunity" where axions are not excluded by astrophysical and cosmological arguments. It is well known that the requirement that stars must not lose energy too efficiently by axions leads to a lower limit on the Peccei-Quinn scale $f_a$ which can be translated into an upper limit on the axion mass by virtue of the relationship $m_a = 0.62\,\mathrm{eV}\,(10^7\,\mathrm{GeV}/f_a)$. It is also well known that the early-universe non-thermal production mechanism leads to an upper bound on $f_a$ (lower bound on $m_a$) lest axions "overclose" the universe. About four years ago several reviews on these topics seemed to be more or less the last word[2]. However, several aspects of these arguments have been modified or refined since and so, it is a good time to review the current situation.

II. STELLAR LIMITS

In analogy to neutral pions, axions generically interact with photons according to $\mathcal{L}_{a\gamma} = g_{a\gamma}\mathbf{E}\cdot\mathbf{B}\,a$ with the coupling constant $g_{a\gamma} = (\alpha/2\pi f_a)\,(E/N - 1.92 \pm 0.08)$. Here, the parameter $E/N$ is a model-dependent fraction of small integers. This coupling allows for the axion decay $a \to 2\gamma$ as well as for the Primakoff conversion $a \leftrightarrow \gamma$ in the presence of external electric or magnetic fields. Because charged particles and photons are abundant in the interior of stars, the hot plasma is an efficient source for the emission of axions. The emission rate with a proper inclusion of screening effects was calculated by virtue of a simple kinetic treatment[3]. Recently, it was calculated in the framework of field theory at finite temperature and density based on the picture that the fluctuating $\mathbf{E}\cdot\mathbf{B}$ term in the plasma is a source for the axion field[4]. In the relevant limit of a classical plasma both calculations yield the same result.

The Primakoff-produced solar axion flux may be searched for with an "axion helioscope" where an x-ray detector looks at the Sun through a long dipole magnet[5]. A first experiment has yielded a negative result[6]. A far more ambitious effort is under way in Novosibirsk where an accelerator dipole magnet was gimballed such that it can follow the Sun; first results are expected later this year[7].

Unfortunately, such efforts are likely in vain because the backreaction of stellar axion emission allows one to derive very restrictive limits on $g_{a\gamma}$. A novel energy-loss mechanism would accelerate the consumption of nuclear fuel in stars and thus lead to the shortening of stellar lifetimes. A case where axion emission would be efficient and the stellar lifetime is well established are low-mass helium-burning stars, so-called horizontal-branch (HB) stars. Low-mass red giants have a degenerate helium core ($\rho \approx 10^6$ g cm$^{-3}$, $T \approx 10^8$ K) so that axion emission is strongly suppressed relative to the cores of HB stars ($\rho \approx 10^4$ g cm$^{-3}$, $T \approx 10^8$ K) whence the number ratio of these stars in globular clusters is a sensitive measure for the operation of axionic energy losses. The observed number ratios agree with standard theoretical expectations to within a few tens of percent so that one finds a limit[8] $g_{a\gamma} \lesssim 0.6 \times 10^{-10}$ GeV$^{-1}$. In GUT axion models where $E/N = 8/3$ this yields $m_a \lesssim 0.4$ eV. The previous "red giant limit" is less restrictive because it was based on the statistically less significant determination of the helium-burning lifetime of the "clump giants" in open clusters[9].

In certain models, axions couple to electrons by $\mathcal{L}_{ae} = (C_e/2f_a)\overline{\psi}_e\gamma_\mu\gamma_5\psi_e\partial^\mu a$ with $C_e$ a model-dependent factor of order unity. For most purposes this derivative coupling is equivalent to the pseudoscalar structure $\mathcal{L}_{ae} = -ig_{ae}\overline{\psi}_e\gamma_5\psi_e a$ with the Yukawa coupling $g_{ae} = C_e m_e/f_a$. In the DFSZ model, $C_e = \frac{1}{3}\cos^2\beta$ with $\beta$ an arbitrary angle. It has recently been suggested that axion emission with $g_{ae} \approx 2\times 10^{-13}$ might dominate the cooling of white dwarfs such as the ZZ Ceti star G117–B15A for which the cooling speed has been established by a direct measurement of the decrease of its pulsation period[10]. Because of this suggestion, a new bound on $g_{ae}$ was derived by a method similar to the above number counts in globular clusters[11]. The resulting limit $g_{ae} \lesssim 3\times 10^{-13}$ or $m_a \lesssim 0.9\times 10^{-2}$ eV$/\cos^2\beta$ is the currently best bound on the axion-electron coupling, but it does not quite exclude the possibility that axions could play a certain role in white dwarf cooling.

These bounds are summarized in Fig. 1. For $m_a$ in excess of typical temperatures in the interior of HB stars and red giants ($T \approx 10$ keV) the axion production would be suppressed. However, the bounds likely reach to larger masses than indicated in Fig. 1 as axions interact quite "strongly" for such large masses and so, they would contribute significantly to the transfer of energy even if their production is substantially Boltzmann-suppressed.

The limits on the axion-photon and axion-electron couplings have been improved somewhat over the past year by a more careful execution of the previous arguments. For the axion-nucleon coupling the trend goes in the opposite direction. The most significant limit is from the cooling speed of nascent neutron stars as established by the duration of the neutrino signal from the supernova (SN) 1987A. Apart from the well-known overall uncertainty of any argument based on only 20 observed events there is an additional problem related to calculations of the axion emission rate from the hot and dense nuclear medium that is believed to exist in the interior of a SN core after collapse.

The axion-nucleon coupling is of the axial-vector type $\mathcal{L}_{aN} = (C_N/2f_a) \bar{\psi}_N \gamma_\mu \gamma_5 \psi_N \partial^\mu a$ so that in the relevant nonrelativistic limit the axion momentum couples to the nucleon spin. These spins fluctuate fast due to the nucleon-nucleon interaction by a spin-dependent force and so, axions are efficiently produced. In a naive perturbative picture the emission rate is computed on the basis of a bremsstrahlung amplitude $NN \to NNa$ with the nucleons interacting by pion exchange. The resulting spin fluctuation rate, however, is found to be so large that one would expect the axion emission rate to be suppressed by destructive interference effects between the axions emitted in subsequent kicks of the nucleon spin. Moreover, the neutrino opacities also depend mostly on an axial-vector coupling. Fast spin-fluctuations would lead to an averaging effect so that neutrinos would no longer "see" a spin and hence could leave the star faster than expected in standard calculations which ignored spin fluctuations entirely. Clearly, neutrino opacities and axion emissivities should be based on a consistent form of the spin-density structure function of the nuclear medium[12].

A recent numerical study of neutron star cooling with modified opacities yields agreement with the SN 1987A signal duration only if the opacities are not suppressed very much[13]. This indicates that the spin fluctuation rate should be far lower than expected in the naive bremsstrahlung picture. The study of Ref. 13 indicates that for the conditions of a SN core ($T \approx 30\,\text{MeV}$) the bremsstrahlung rate should saturate at about 10% nuclear density, leading to a very approximate bound of $m_a \lesssim 10^{-2}\,\text{eV}$. This is about an order of magnitude less restrictive than had been thought previously.

If axions interact too strongly they are trapped and contribute to the transfer of energy rather than to a direct cooling of the inner SN core. Axions with $m_a \gtrsim 10\,\text{eV}$ probably cannot be excluded on the basis of the duration of the SN 1987A neutrino signal[14].

However, axions with masses larger than this, i.e., with stronger interactions, could actually cause a significant contribution to the signal measured at the IMB and Kamiokande II water Cherenkov detectors by their absorption on $^{16}$O and the subsequent emission of $\gamma$ rays. To avoid too many events one can exclude the range[15] $20\,\text{eV} \lesssim m_a \lesssim 20\,\text{keV}$.

## III. COSMOLOGICAL LIMITS

If axions were sufficiently strongly interacting ($f_a \lesssim 10^8$ GeV) they would have come into thermal equilibrium before the QCD phase transition and so, we would have a background sea of invisible axions in analogy to the one expected for neutrinos[16]. This parameter range is excluded by the astrophysical arguments summarized in Fig. 1 and so, axions must be so weakly interacting that they have never come into thermal equilibrium. Still, the well-known misalignment mechanism will excite coherent oscillations of the axion field[17]: When the temperature of the universe falls below $f_a$ the axion field settles somewhere in the brim of its Mexican hat potential. When the hat tilts at the QCD phase transition, corresponding to the appearance of a mass term for the axion, the field begins to move and finally oscillates when the expansion rate of the universe has become smaller than the axion mass. In units of the cosmic critical density one finds for the axionic mass density

$$\Omega_a h^2 \approx 0.23 \times 10^{\pm 0.6} (f_a/10^{12}\,\text{GeV})^{1.175}\, \Theta_i^2\, F(\Theta_i) \qquad (1)$$

where $h$ is the present-day Hubble expansion parameter in units of $100\,\text{km}\,\text{s}^{-1}\,\text{Mpc}^{-1}$. The stated range reflects recognized uncertainties of the cosmic conditions at the QCD phase transition and uncertainties in the calculations of the temperature-dependent axion mass. The cosmic axion density thus depends on the initial misalignment angle $\Theta_i$ which could have any value in the range $0 - \pi$. The function $F(\Theta_i)$ encapsules anharmonic corrections to the axion potential for $\Theta \gg 0$. (For a recent analytic determination of $F$ see Ref. 18.)

The age of the universe indicates that $\Omega h^2 \approx 0.3$, causing a problem with $\Omega = 1$ models if $h$ is around 0.8 as indicated by recent measurements. For the present purpose I take $\Omega_a h^2 = 0.3 \times 2^{\pm 1}$ for axions which constitute the dark matter where the adopted uncertainty of a factor of 2 likely covers the whole range of plausible cosmological models. Then, axions with $m_a = \mathcal{O}(1\,\mu\text{eV})$ are the cosmic dark matter if $\Theta_i$ is of order 1.

Because the corresponding Peccei-Quinn scale of $f_a = \mathcal{O}(10^{12}\,\text{GeV})$ is far below the GUT scale one may speculate that cosmic inflation, if it occurred at all, did not occur after the PQ phase transition. If it did not occur at all, or if it did occur before the PQ transition with $T_{\text{reheat}} > f_a$, the axion field will start with a different $\Theta_i$ in each region which is causally connected at $T \approx f_a$ and so, one has to average over all possibilities to obtain the present-day axion density. More importantly, because axions are the Nambu-Goldstone mode of a complex Higgs field after the spontaneous breaking of a global U(1) symmetry, cosmic axion strings will form by the Kibble mechanism[19]. The motion of these global strings is damped primarily by the emission of axions rather than by gravitational waves. At the QCD phase transition, the U(1) symmetry is explicitly broken (axions acquire a mass) and so, domain walls bounded by strings will form, get sliced up by the interaction with strings, and the whole string and domain wall system will quickly decay into axions. This complicated sequence of events leads to the production of the dominant contribution of cosmic axions. Most of them are produced near the QCD transition at $T \approx \Lambda_{\text{QCD}} \approx 200\,\text{MeV}$. After they acquire a mass they are nonrelativistic or mildly relativistic so that they are quickly redshifted to nonrelativistic velocities. Thus, even the string and domain-wall produced axions form a cold dark matter component.

In their recent treatment of axion radiation from global strings, Battye and Shellard[20] found that the dominant source of axion radiation are string loops rather than long strings, contrary to what was assumed in the previous works by Davis[19] and Davis and Shellard[21]. At a given cosmic time $t$ the average loop creation size is parametrized as $\langle\ell\rangle = \alpha t$ while the radiation power from loops is $P = \kappa\mu$ with $\mu$ the renormalized string tension. The exact values of the parameters $\alpha$ and $\kappa$ are not known; the cosmic axion density is a function of the combination $\alpha/\kappa$. For $\alpha/\kappa < 1$ the dependence of $\Omega_a h^2$ on $\alpha/\kappa$ is found to be rather weak. Battye and Shellard favor $\alpha/\kappa \approx 0.1$ for which $\Omega_a h^2 = 18 \times 10^{\pm 0.6}(f_a/10^{12}\,\text{GeV})^{1.175}$, about an order of magnitude smaller than originally found by Davis and Shellard[19,21]. The overall uncertainty has the same source as in Eq. (1) above. With $\Omega_a h^2 = 0.3 \times 2^{\pm 1}$ the mass of dark-matter axions is found to be $m_a = 30 - 1000\,\mu\text{eV}$; the cosmologically excluded range of axion masses is indicated in Fig. 1.

These results are plagued with systematic uncertainties. Battye and Shellard argue that the largest uncertainty was the impact of the backreaction of axion emission on the string network. (They believe that a current numerical study[22] will allow them to pin down the parameter $\alpha/\kappa$ to within, say, a factor of 2.) Further, Sikivie and his collaborators[23] have consistently argued that the motion of global strings was overdamped, leading to an axion spectrum emitted from strings or loops with a flat frequency spectrum. In Battye and Shellard's treatment, wavelengths corresponding to the loop size are strongly peaked, the motion is not overdamped. In Sikivie et al.'s picture, much more of the string-radiated energy goes into kinetic axion energy which is redshifted so that ultimately there are fewer axions; it was argued that the cosmic axion density was then of order the misalignment contribution. Therefore, following Sikivie et al. one would estimate the mass of dark-matter axions at about $m_a = 4 - 150\,\mu\text{eV}$ where the range reflects the same overall uncertainties that bedevil the Battye and Shellard estimate, or the misalignment contribution.

While the cosmic axion bounds claimed by both groups of authors still differ significantly, the overall uncertainty within either scenario is larger than the mutual disagreement, i.e., the range of masses where axions could be the dark matter overlaps significantly between the predictions of the two groups (Fig. 1). Moreover, there remain difficult to control uncertainties, for example, with the "dilute instanton gas" calculation of the temperature-dependent axion mass near the QCD phase transition. There may be other unaccounted systematic problems which may increase the adopted uncertainty of the cosmic mass prediction which is represented in Fig. 1 by the slanted end of the cosmic exclusion bar. Still, it would be of great importance to resolve the disagreement between the two groups of authors once and for all.

IV. PHASE SPACE DISTRIBUTION

Even though axions play the role of a cold dark matter candidate, their phase-space distribution in the galaxy may well exhibit novel features which are of relevance for the search experiments. Hogan and Rees[24] have pointed out that the different initial misalignment angles of the axion field in different causally connected regions lead to density fluctuations which are nonlinear from the very beginning of matter domination in the early universe. This leads to the formation of "axion mini clusters" which may partially survive galaxy formation and thus can be found in the Milky Way today. For a suitably large initial density contrast these clusters can condense into axionic boson stars by virtue of higher-order axion-axion couplings[25]. The collision rate of mini clusters with the solar system would be low so that the direct search experiments need sufficient sensitivity to pick up the diffuse component of the galactic axions. While it appears unlikely that a large fraction of the galactic axions is locked up in mini clusters this is an issue that may deserve further study.

Another interesting possibility is that axions may have maintained some of their initial phase-space distribution, i.e., that they are not fully virialized in the galaxy. Then, the axion velocity distribution would exhibit very narrow peaks[26] which ultimately could be studied in the laboratory. The velocity distribution of galactic axions is currently under further study[27].

V. SUMMARY

The astrophysical and cosmological limits on axions leave a narrow window of parameters where axions could still exist (Fig. 1); they would then be a significant fraction or all of the dark matter of the universe. While the SN 1987A as well as the cosmological bound are each very uncertain, there is a recent trend toward an allowed range near $m_a = \mathcal{O}(1\,\mathrm{meV})$ where no current or proposed experimental effort appears to be sensitive. Of course, the overall quantitative uncertainty of the predicted cosmic axion density is large, especially if one includes the possibility of late-time inflation after the Peccei-Quinn phase transition or late-time entropy production after the QCD phase transition. Therefore, the microwave cavity experiments in Livermore and Kyoto no doubt have a fair chance of detecting galactic axions if they are the dark matter. Still, suggestions for a meV-axion detector are hugely welcome!

ACKNOWLEDGEMENTS

For the preparation of this contribution I have greatly benefitted from conversations or electronic communications with R. Battye, K. van Bibber, M. Khlopov, P. Sikivie, and D. Spergel.

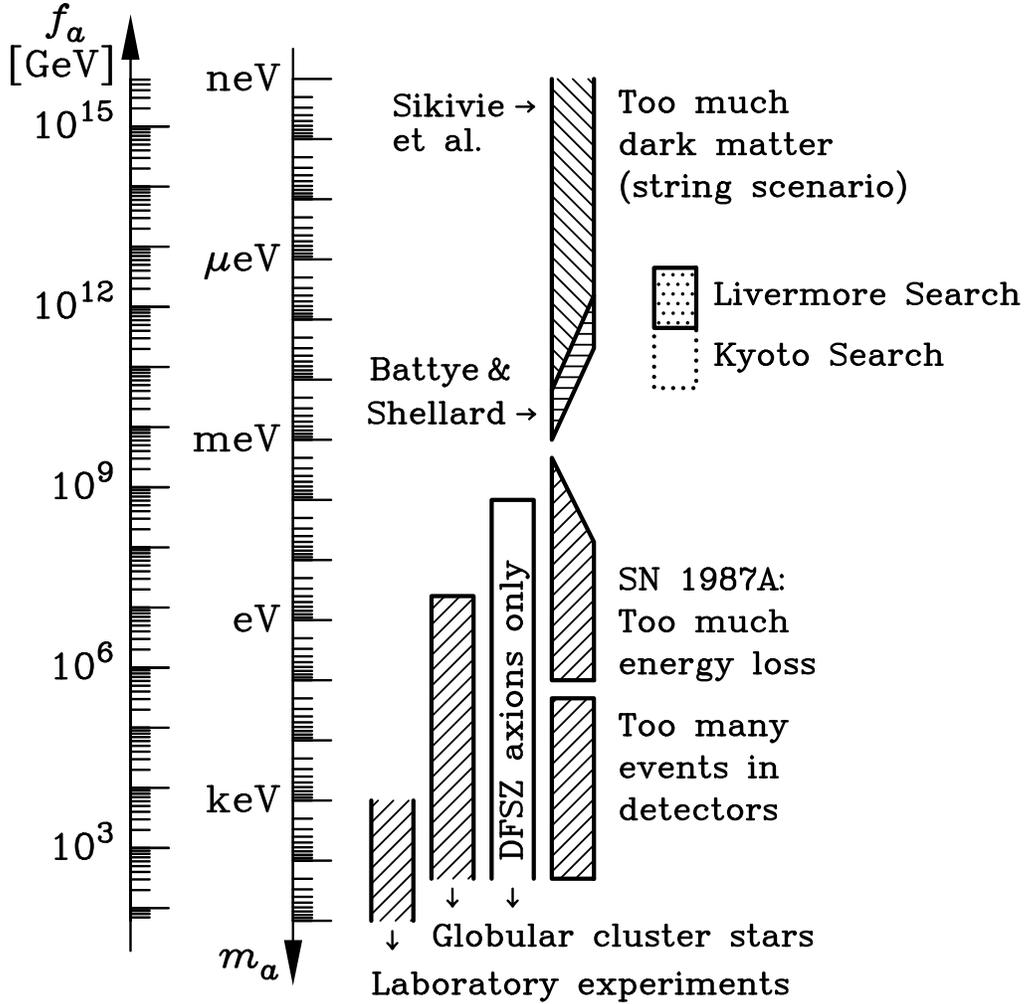

Figure 1: Summary of astrophysical and cosmological axion bounds as described in the text. The hatched exclusion regime from globular cluster stars assumes an axion-photon coupling corresponding to $E/N = 8/3$ as in GUT models. The bound "DFSZ axions only" is based on an axion-electron coupling with $\cos^2\beta = 1$. The uncertainty of the SN 1987A cooling limit, indicated by the sloping end of the exclusion bar, is only a crude estimate. The Livermore and Kyoto Search experiments are discussed by K. van Bibber and S. Matsuki in their contributions to these proceedings.